\documentclass[preprint,twocolumn]{aastex631}
\usepackage{xcolor}

\usepackage{xcolor}

\newcommand{\thickbar}[1]{\mathbf{\bar{\text{$#1$}}}}
\shorttitle{Recombination of hot ionized nebulae: V4334 Sgr}
\shortauthors{Reichel et al.}
\graphicspath{{./}{figures/}}
\usepackage{amsmath}
\begin{document}

\title{Recombination of hot ionized nebulae:
   The old planetary nebula around V4334 Sgr (Sakurai's star)\footnote{This investigation makes use of ESO data from program IDs \mbox{077.D-0394}, \mbox{079.D-0256}, \mbox{381.D-0117}, \mbox{383.D-0427}, \mbox{385.D-0292}, \mbox{087.D-0223}, \mbox{089.D-0080}, \mbox{091.D-0209}, \mbox{093.D-0195}, \mbox{095.D-0113}, \mbox{097.D-0146} and \mbox{099.D-0045}.}
   }

\correspondingauthor{Stefan Kimeswenger}
\email{Stefan.Kimeswenger@uibk.ac.at}

   \author[0000-0002-1365-9336]{Martin Reichel}%\inst{1}
    \affiliation{Institut f{\"u}r Astro- und Teilchenphysik, Universit{\"a}t Innsbruck, Technikerstr. 25\/8, 6020 Innsbruck, Austria}
   \author[0000-0003-2379-0474]{Stefan Kimeswenger}%\inst{1,2}%,\thanks{Corresponding Author}}
    \affiliation{Institut f{\"u}r Astro- und Teilchenphysik, Universit{\"a}t Innsbruck, Technikerstr. 25\/8, 6020 Innsbruck, Austria}
    \affiliation{Instituto de Astronom{\'i}a, Universidad Cat\'olica del Norte, Av. Angamos 0610, Antofagasta, Chile}    
       \author[0000-0001-7490-0739]{Peter A.M. van Hoof}%\inst{3}
\affiliation{Royal Observatory of Belgium, Ringlaan 3, B-1180 Brussels, Belgium}
        \author[0000-0002-3171-5469]{Albert A. Zijlstra}%\inst{4,5}
\affiliation{Jodrell Bank Centre for Astrophysics, Alan Turing Building, University of Manchester, Manchester, M13 9PL, UK}
%\affiliation{Laboratory for Space Research, The University of Hong Kong, Pokfulam Road, Hong Kong}

         %\andKIMESWENGER/ VAN HOOF/ BARIA/ ZIJLSTRA/ HAJDUK/ HERWIG/ VAN DE STEENE/ KAY
        \author[0000-0003-4754-4673]{Daniela Barr{\'i}a}%\inst{2}
    \affiliation{Facultad de Ingeniería y Arquitectura, Universidad Central de Chile, Av. Francisco de Aguirre 0405, La Serena, Coquimbo, Chile}

        \author[0000-0001-6028-9932]{Marcin Hajduk}%\inst{6}
\affiliation{Space Radio-Diagnostics Research Centre, University of Warmia and Mazury, Prawoche{\'n}skiego 9, 10-720 Olsztyn, Poland}

        \author[0000-0001-7628-7499]{Griet C. Van de Steene}%\inst{3}
\affiliation{Royal Observatory of Belgium, Ringlaan 3, B-1180 Brussels, Belgium}

\author[0000-0002-2149-2660]{Daniel Tafoya}
\affiliation{Department of Space, Earth and Environment, Chalmers University of Technology, \\
	Onsala Space Observatory, 439~92 Onsala, Sweden}



\begin{abstract}

%   Recombination of ionized nebulae at low densities cannot be observed normally, without suffering from a strong influence from the evolutionary time scale of the ionizing source. 
%   Thus we were lacking an isolated observational verification and possibility to  improve models of recombination processes in very thin plasma.  Laboratory experiments are unable to reach appropriate conditions. 
%   However, the outstanding case of the extended nebula around the very late helium flash (VLTP) star V4334 Sgr is giving us a unique laboratory for this kind of study. 
  After becoming ionized, low density astrophysical plasmas will begin a process of slow recombination. Models for this still have significant uncertainties. The recombination cannot normally be observed in isolation, because the ionization follows the evolutionary time scale of the ionizing source. Laboratory experiments are unable to reach the appropriate conditions because of the required very long time scales.
 The extended nebula around the very late helium flash (VLTP) star V4334 Sgr provides a unique laboratory for this kind of study.    The sudden loss of the ionizing UV radiation after the VLTP event has allowed the nebula to recombine free from other influences. More than 290 long slit spectra taken with FORS1/2 at the ESO VLT between 2007 and 2022 are used to follow the time evolution of lines of H, He, N, S, O, Ar.
%   \ion{H}{1}, \ion{He}{1}, \ion{He}{2}, $[$\ion{N}{2}$]$, $[$\ion{S}{2}$]$, $[$\ion{S}{3}$]$, $[$\ion{O}{3}$]$ and $[$\ion{Ar}{3}$]$.  
Hydrogen and helium lines, representing most of the ionized mass, do not show significant changes. A small  
   increase is seen in [\ion{N}{2}] (+2.8\,\% yr$^{-1}$; significance 2.7 $\sigma$), while we see a decrease in [\ion{O}{3}] (-1.96\,\% yr$^{-1}$; 2.0 $\sigma$). The [\ion{S}{2}] lines show a change of +3.0\,\% yr$^{-1}$; 1.6 $\sigma$).
   The lines of [\ion{S}{3}] and of [\ion{Ar}{3}] show no significant change. For [\ion{S}{3}], the measurement differs from the predicted decrease by 4.5$\sigma$.  A possible explanation is that the fraction of  [\ion{S}{4}] and higher is larger than expected. 
   Such an effect could provide a potential solution for the sulfur anomaly in planetary nebulae.

\end{abstract}



%
%-------------------------------------------------------------------
\section{Introduction} \label{sec:intro}

When the UV radiation which photo-ionizes a nebula ceases, the ions begin to recombine. This is an important process 
in photoionized nebulae with typical densities of a few tens to a few hundred particles cm$^{-3}$, where the slow recombination leaves the nebula out of equilibrium.  These conditions exist in low density \ion{H}{2} regions, in planetary nebulae and in nebulae around post common envelope binaries of the BE\ UMa family.  Recombination at these low densities is not well studied, neither in astrophysical environments nor in the laboratory.

Laboratory studies aim to measure the effects directly. But  appropriate density and excitation conditions are difficult to reach. The leading edge experiment Cryogenic Storage Ring (CSR) has recently reached particle densities down to 140 cm$^{-3}$  \citep{CSR19}. In order to keep the free flight time (avoiding wall interaction) larger than the  ion lifetimes, temperatures of a few Kelvin are used. This is suitable for simulating conditions of cold molecular clouds in the  interstellar medium (ISM). However, these have very low ionization levels.  Ionized nebulae have typical temperatures of $T_{\rm e} \gtrapprox 10\,000\,\,$K. As then the mean free path of particles reaches thousands of kilometers, chamber wall interaction would dominate laboratory experiments.

%Mono-atomic gases containing simple H-like atoms are well modelled purely with the transition probabilities \citep[see e.g. chapter 2 in][]{OF06}. However, these authors already conclude that simple transfer of angular momentum in case of the He$^{++}$ zone complicates the calculations. Observational and calculated recombination factors may differ significantly. % by 3 orders of magnitude, e.g.  for H$_3^+$ \citet{H3plus}.

Astrophysical targets are needed for the determination of the recombination timescales under these conditions, and for the verification of theoretical models. While ionization happens on timescales of the light travel time through the nebulae and thus within weeks or months, recombination happens very slowly \citep{Schoenberner08, Balick21}.  The latter authors estimate that the expected recombination rate of the mono-atomic gas per
ion is 
\begin{equation}
    R_{\rm rec} = 10^5 {\rm yr}/(Z^2 n_{\rm e}){\rm ,}
    \label{equ:01}
\end{equation}
 
\noindent where $Z$ is the ionic
charge and $n_{\rm e}$ is the electron density. 
These calculations are based on a mono-atomic gas in a two-level approximation \citep{OF06}  and assume that the ion is in the ground state.
The latter assumption holds only for H, He and noble gases, where the first excitation level is at $>$ 10\,eV and thus collisional excitation is not contributing at all. Once fluorescence and interactions between ionic species are taken into account, excitation may change some approximations \citep[see discussion in Sec. 3.2.5 in][]{Cloudy2017}.
The interactions between species are partly included in a simplified model by \citet{Koskela2012} for elements up to oxygen.
%, based on \citet{Binette2003}. 
%However, they did not include a full description of some of the candidate processes like  fluorescence via lines with matching wavelengths.
\citet{Cloudy2017} discuss in the implementation of the full collisional-radiative mode (CRM) of the H- and He-like isoelectronic sequences, that ionization out of highly excited states can alter the ionization balance at some density regimes. Whether this is  an important issue in our targets remains an open question. 
%New developments in the CLOUDY code\footnote{\url{http://www.nublado.org}} will do a more sophisticated handling throughout many processes
%. However, the first pure model results presented are based on fairly dense nebulae only 
%\citep{Cloudy2020}.

For most nebulae, the dominant astrophysical evolutionary timescale is that of their ionizing source. The source (star)  may heat up very rapidly, but, as mentioned, the ionizing timescale is anyhow very short \citep{Schoenberner08}. Cooling or dimming of the ionizing source, causing recombination, happens on timescales of thousands to hundred thousands of years in most targets of interest. An exception might be the  planetary nebulae (PNe)  around massive progenitor stars
%, which can dim on timescales of decades.
However, their nebulae are very dense %in the central regions, 
and only in the very outskirts weak effects can be determined \citep[e.g. NGC\ 6445, ][] {vanHoof2000}.

Hence the recombination evolution is dominated by the dimming/cooling timescale of the ionizing source. 
There are only a handful of exceptional targets, suffering fast cooling and/or dimming, where we in fact are able to investigate isolated recombination in--situ without major obfuscation caused by other processes. Here we present the observational study of \object{V4334 Sgr} spanning a period of more than 16 years. It suffered a very late thermal pulse (VLTP) which caused a sudden loss of the ionizing UV radiation. This caused the onset of recombination in the old PN surrounding V4334 Sgr.
\citet{Pollacco02} finds no discernible recombination through the first 3 years of evolution after the event whilst  \citet{Schoenberner08} argues that the total timescale is dominated by the ionization timescale and that a new recombined equilibrium is reached after only a few months.

%\section{Potential targets and their environment}
\section{The Target}

In 1996 \citet{Nakano1996} discovered the re-brightening of Sakurai's Star (V4334 Sgr) as a cool post-AGB star. It was surmised to have undergone a   very late thermal pulse (VLTP)  around 1991 or 1992 \citep{Herwig2001,Lawlor2003,Hajduk2005, MillerBertolami2006}. The star was found to be at the center of an old planetary nebula. After the eruption, the star expelled a new shell, located inside the old, now recombining PN.  Since then spectra have been taken frequently.
These provide the first closely spaced time series of spectra of an ionized nebula that undergoes recombination without any reionization by UV photons emitted by the central star.

Evolutionary models predict re-heating of the star of V4334 Sgr to temperatures that would result in renewed photoionization of the nebula \citep{Hajduk2005} but they differ on the expected timescale for this. However, no evidence for a hot stellar continuum or stellar wind features have been found in the optical nor infrared spectra so far \citep{Hinkle2020}. No evidence for increasing free-free radio emission from the newly ejected matter  has been found either (Hajduk et al., in prep.). Moreover,  our FORS long slit spectra show a very red continuum of the central source  V4334\ Sgr in the optical.  During the first years of the campaign, the continuum of the central source was not detected. Later it slowly appeared above the detection limit. Stacking the spectra  taken in 2017 lead us to an estimated (B-V) of the continuum of about +3\fm4. If the central source had heated to something hotter than 10\,000\;K, the intrinsic colors -0\fm3$\le$(B-V)$_0 \le$0\fm0 \citep{Ducati01} would require optical extinctions of $A_{\rm V} >$  11\fm5. 
%using the extinction law by \citet{Cardelli}. 
 This assumes that the observed continuum is generated by direct light. Scattering at circumstellar clouds, efficient in the the blue bands, leads, like the  scattered-light residual in R\ CrB stars, to an underestimate of the extinction \citet{RCrB}. \ The value fits well to the extinction derived by the photometric evolution when the shell was formed \citep{Pavlenko01}. Similar results for the extinction were found by \citet{Koller01} for the older twin V605~Aql by comparing optical to infrared luminosity. This extreme extinction inhibits potential UV photons from the reheating central star from reaching the region of the old PN. Our ALMA observations \citep{ALMA1,ALMA2} of HCN and HNC molecules in the newly ejected material give us an estimate of the UV flux in the core of 0.5\,erg~cm$^{-2}$~s$^{-1}$, based on the recent investigation by \citep{HCN_UV} in the clumps of the Helix nebula (NGC~7293). The molecular gas detected by ALMA is at most 0.3\arcsec\ from the core \citep{Hinkle2020}. This gives an upper limit for the total UV flux in the compact core $\le$~0.1~L$_\odot$. Moreover, this radiation will have a geometrical thinning of at least 1/500 before reaching the outer location of the old PN. Re-ionization of the old PN by UV photons from the post-VLTP star thus can currently be neglected for this study.

 Other potential targets which could be used to study recombination have been discussed in the past. These include the late thermal pulse (LTP) stars \object{FG Sge} \citet{Tylenda80, Arkhipova09} and \object{Hen 3-1357} (Stingray Nebula) \citep{Rendl17,Lawlor21,Balick21}, the VLTP star \object{V605 Aql} \citep[studied in][]{Lechner2004, Koskela2012} and the very young dense PN \object{SwSt 1} \citet{SwSt1}. Also, the nebula around the family of post-common-envelope pre-cataclysmic BE UMa variables feature recombination effects \citep[][ and references therein]{Mitrofanova2016,YY_Hya2020}.
However, all those targets do either not have such a rapid evolution of the ionizing UV radiation, have much larger densities and thus very fast recombination in equilibrium, or are lacking an appropriate time coverage with homogeneous spectral data.

Thus our %latter 
target provides a unique laboratory where the UV radiation was suddenly and completely switched off.

\begin{figure}[t]
\centering{\includegraphics[width=0.46\textwidth]{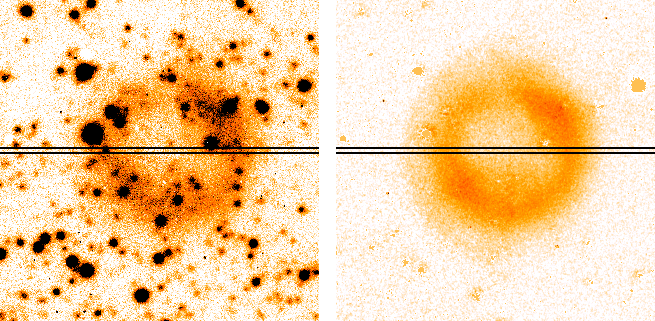}}
\caption{ The old PN around V4334 Sgr. Left: a single 300 seconds frame taken with an [\ion{O}{3}] filter; Right: after star removal and averaging { the two available} frames (see text). The black lines indicate the position of our slit. \ The old PN is 44\arcsec across. }\label{fig:image}
\end{figure}
%\newpage

\section{Data}

 The data sample used consists of a set of a few hundred long slit spectra crossing the old PN nebula around the centraL VLTP object V4334 Sgr.\ 
These were obtained with the Very Large Telescope (VLT) unit telescopes (UTs) 1 and 2 using the FORS1/2 spectrographs \citep{Appenzeller1998}. The original purpose of the observations was to study the changes in the central star V4334 Sgr itself. Since the slit was always aligned in the same direction it is also possible to study the outer nebula as an unintended but perfectly homogeneous byproduct. The entire project was conducted in service mode demanding good weather conditions (CLR).  However, during the data reduction, some spectra with lower quality showed up. The data originally covered 11 observational epochs from 2007 to 2017 with 265 spectra. A few spectra were removed from this study, as the slit centering was not perfect in one run in 2011 \citep{Martin}. In 2022 (3$^{\rm rd}$ to 10$^{\rm th}$ of April) an additional set of 18 spectra were taken, improving the timeline which now spans over 16 observing seasons.
  The service mode logs always mention at least clear weather conditions and good seeing. However, after the additional calibration processing, we rejected 46 spectra due to poor seeing or significant zero point offsets. Thus a total sample of 237 spectra remained.

The VLT observations used five  different low resolution GRISMs\footnote{FORS VLT-MAN-ESO-13100-1543 Issue 101, 24/08/2017:\protect{\newline} \url{http://www.eso.org/sci/facilities/paranal/instruments/fors/doc/VLT-MAN-ESO-13100-1543_P01.pdf}} \citep{FORS, Martin}. Table~\ref{GRISM} provides an overview of the covered wavelength regime and the resolution near the central wavelength. There is some overlap between the spectra from the red and the blue setup which can be used as an additional test of the calibration quality and homogeneity of the data.

The data were taken between April and August of each year, with an East-West orientated slit.  Two images, obtained from the ESO archive, were taken with 
FORS1 October 2$^{\rm nd}$ 2002 using an [\ion{O}{3}] narrow band filter. After each frame a second frame was taken using a filter for extragalactic observations redshifted by 6000\,km\,s$^{-1}$. Those images were used to remove the stars. The slit position is indicated in Fig. \ref{fig:image}.  A slit width of 1\farcs0 was used until 2016. In 2017 and 2022 a 1\farcs3 slit was used to cover the new ejecta of  the VLTP which were growing in size.  Further data was taken with FORS by other groups covering the time just after the outburst. However, as slit position and orientation varied and mostly were far from our target position, they are not used here.

\begin{table}[b]
  \centering
  \caption{The GRISM configurations that were used according to \citet{FORS}. However, the usable wavelength range is given, ignoring the unused badly illuminated edges of the CCD.  $N$ gives the number of frames in the final sample with each grism.}
    \begin{tabular}{ccccc}
 \hline\hline 
%  & Wavelength & & \\
  &  range &  resolution & dispersion & $N$ \\
 GRISM & (\AA)  & $\lambda/\Delta \lambda$& (\AA/pixel) &\\
    \hline
    600V & 4600--7096 & 990   & 0.74 &  23 \\
    600I & 6790--8880 & 1500  & 0.66 &   12 \\
    600z & 7466--10153 & 1260  & 0.81 &   3\\
    300V & 4610--8590 & 440   & 1.68 &   99 \\
    300I & 6150--10190 & 660   & 1.62 &   100 \\
    \hline
    \end{tabular}%
  \label{GRISM}%
  \end{table}%

Observations were always conducted with the same sequence centering an offset star in direct images and blind-offsetting to the faint target. Three spectra with exposure times around 900 seconds were taken. Such a sequence forms one observing block (OB).
If another OB followed directly, the centering was redone  to insure against pointing drifts or flexure. For each year and each GRISM two to four of these OBs were taken. 
The wavelength and basic flux calibration was derived by the ESO FORS 
pipeline\footnote{VLT-MAN-ESO-19500-4106 vers. 5.12, June 2020:\protect{\newline} \url{https://ftp.eso.org/pub/dfs/pipelines/instruments/fors/fors-pipeline-manual-5.12.pdf}}
\citep[ver. 5.6.2;][]{fors_pip} with the night sky removal deactivated to avoid partial removal of nebular lines. The further manual steps were obtained in the framework of ESO MIDAS \citep{MIDAS,MIDAS_1,MIDAS_2}.  These manual steps were a) remove sky emission lines, b) correct for sky transparency variations, c) correct for seeing slitloss, and d) manual extraction of the PN emission lines from the 2D spectra. Small MIDAS batch files were written to speed up and homogenize this interactive process. The four steps are described in detail below. \ 

 As the ESO pipeline is dedicated to stellar spectroscopy, the automated search for night sky emission lines also removes partially the lines of extended emission line objects like PNe. Thus this process was disabled in the pipeline. Moreover, along the 7 arcmins long slit some remnant of the optical curvature remains after the pipeline reduction. Star-free regions along the slit just outside the PN only about 20 arcsec from the nebula were used to derive a night-sky emission line spectrum which was used to subtract from the region of the PN in the 2D spectra.  

\begin{figure}
     \centering
     \includegraphics[width=0.46\textwidth]{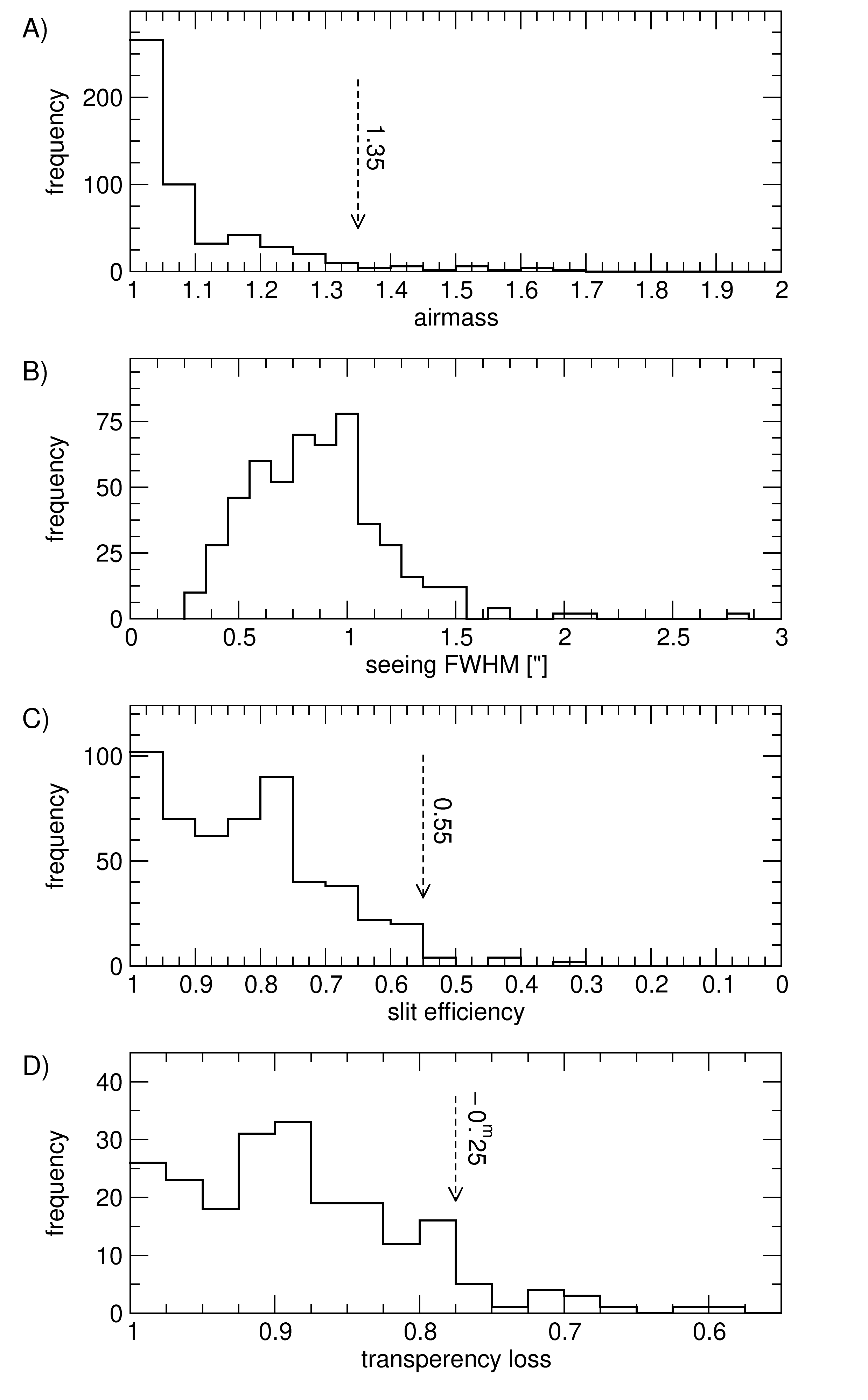}
     \caption{frequency distribution of the A) airmass, B) seeing, C) from the seeing calculated slit efficiency and D) measured transparency losses. The arrows indicate the rejection limits applied for the sample (see text).}
     \label{fig:histogramms}
\end{figure}

 The standard extinction curve derived by \citet{Patat2011} is automatically applied by the ESO pipeline. The typical airmass was between 1 and 1.1 (see Fig \ref{fig:histogramms}A). Thus the corrections were small.  However as we had shown in observations of the scattered light at several nights near full moon some offsets of the transparency can occur even in very good nights \citep{Jones2013,Jones2019}. The ESO standard stars, automatically used by the pipeline, are taken not very frequently during nights in the service mode calibration plan.  Only a few of our OBs were taken at higher airmasses  $>$ 1.35.  Those where rejected from the final sample.

 %\\
%Unfortunately not all recorded images were useful. In 2011 with the 300V GRISM the target was missed. %Hence the observed region of the PN had changed endangering homogeneity. 
%In 2016 no images with the 300V GRISM were taken. %(Fig.~\ref{fig:calibration_stars}). 
 
% \begin{figure*}
%     \centering
%     \includegraphics[width=\textwidth]{slit_reference.png}
%     \caption{The selection for calibration stars. Upper panel: direct image taken through the slit without the GRISM, expanded in the $y$ direction to enhance the visibility of the centering. Thus the stars do not appear round. The blue box marks the size of a single pixel in this direction. The FWHM of the stellar images in the $x$ direction was 2.8 pixels. The selected stars are marked with the red bars. Lower panel: A fraction of the spectrum (before distortion correction)  directly after the upper slit centering image by adding the dispersion GRISM only.}
%     \label{fig:calibration_stars}
% \end{figure*}

 \begin{figure*}
     \centering
     \includegraphics[width=\textwidth]{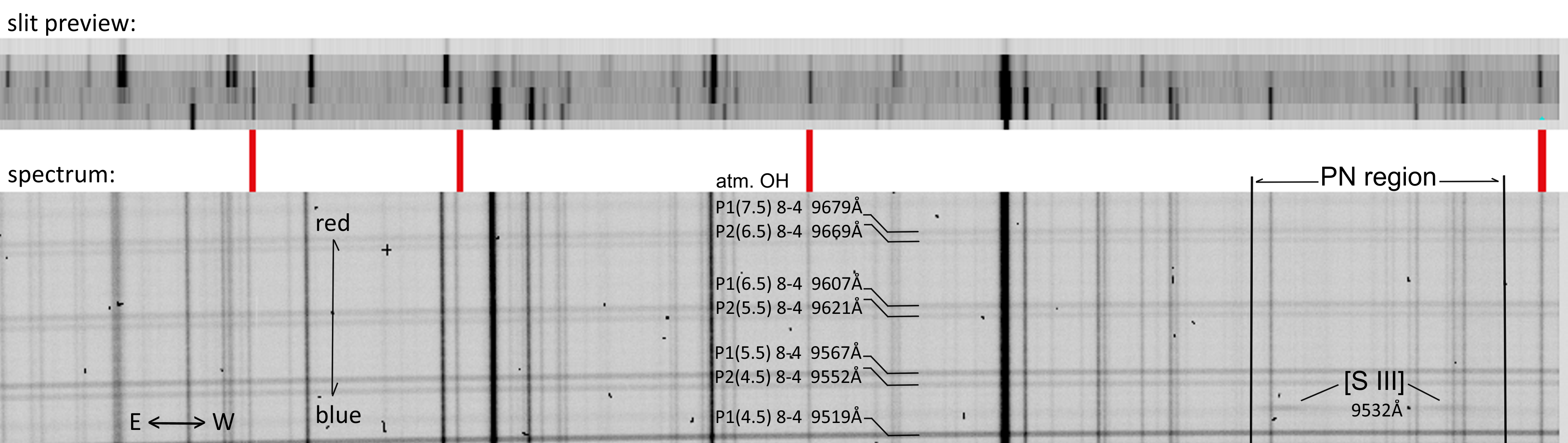}
     \caption{ The selection for calibration stars. { The upper panel shows t}he direct image taken through the slit (expanded in the $y$ direction to enhance the visibility of the centering) { before adding the GRISM. Only well centered stars were selected for calibration purposes. The lower panel shows a} fraction of the red end of the spectrum taken directly thereafter.
     %Thus the stars do not appear round. The blue box marks the size of a single pixel in this direction. The FWHM of the stellar images in the $x$ direction was 2.8 pixels. 
     The selected calibration stars are marked with the red bars { connecting their image in the slit preview with their appearance in the spectrum. The night sky line identifications are taken from \citet{OH_night_lines}.} %Lower panel: A fraction at the red end of the spectrum (before distortion correction) taken directly after the upper slit image. Wavelength direction is blue down to red up, spatial direction along the slit is east left and west right.
     }
     \label{fig:calibration_stars}
\end{figure*}

    The  flux\ zero points then were checked by the flux of field stars well centered in the slit.  At the begin of an OB a slit preview image without the grism is taken for centering purposes. This was used to identify stars along the slit which are well centered. { Solely those give a good calibration factor for the slitloss correction depending on the seeing only.} Six stars (four on chip one which also contains the optical center of the FOV and thus the PN and two on the second chip) were identified (Fig. \ref{fig:calibration_stars}). Six frames from one night in 2011 with GRISM 300V were rejected as the sky region was taken erroneously with a few arcseconds offset. The ESO Ambient Conditions Data Base (ASM)\footnote{\url{http://archive.eso.org/cms/eso-data/ambient-conditions.html}} shows for the best photometric nights the zero points given in \citet{Patat2011}. Those 26 spectra are used as the basis for the transparency 1.0.
    In fact, we measure later only relative variations with time. Therefore, the chosen absolute zeropoint is not of importance here. The median of the flux, integrated along the whole spectrum of the field stars, is used to define the offset. The scatter among them was of the order of $0\fm05$. Thus the differential error of the calibration of individual spectra should be in the order of  ${6}^{-0.5} \times 0\fm05 = 0\fm02$. Figure \ref{fig:histogramms}D) shows the distribution of the derived transparency values we used for corrections. They correspond well to the values of the ASM. We set a rejection limit at a transparency loss of -0\fm25.

\begin{figure}
\centering{\includegraphics[width=0.40\textwidth]{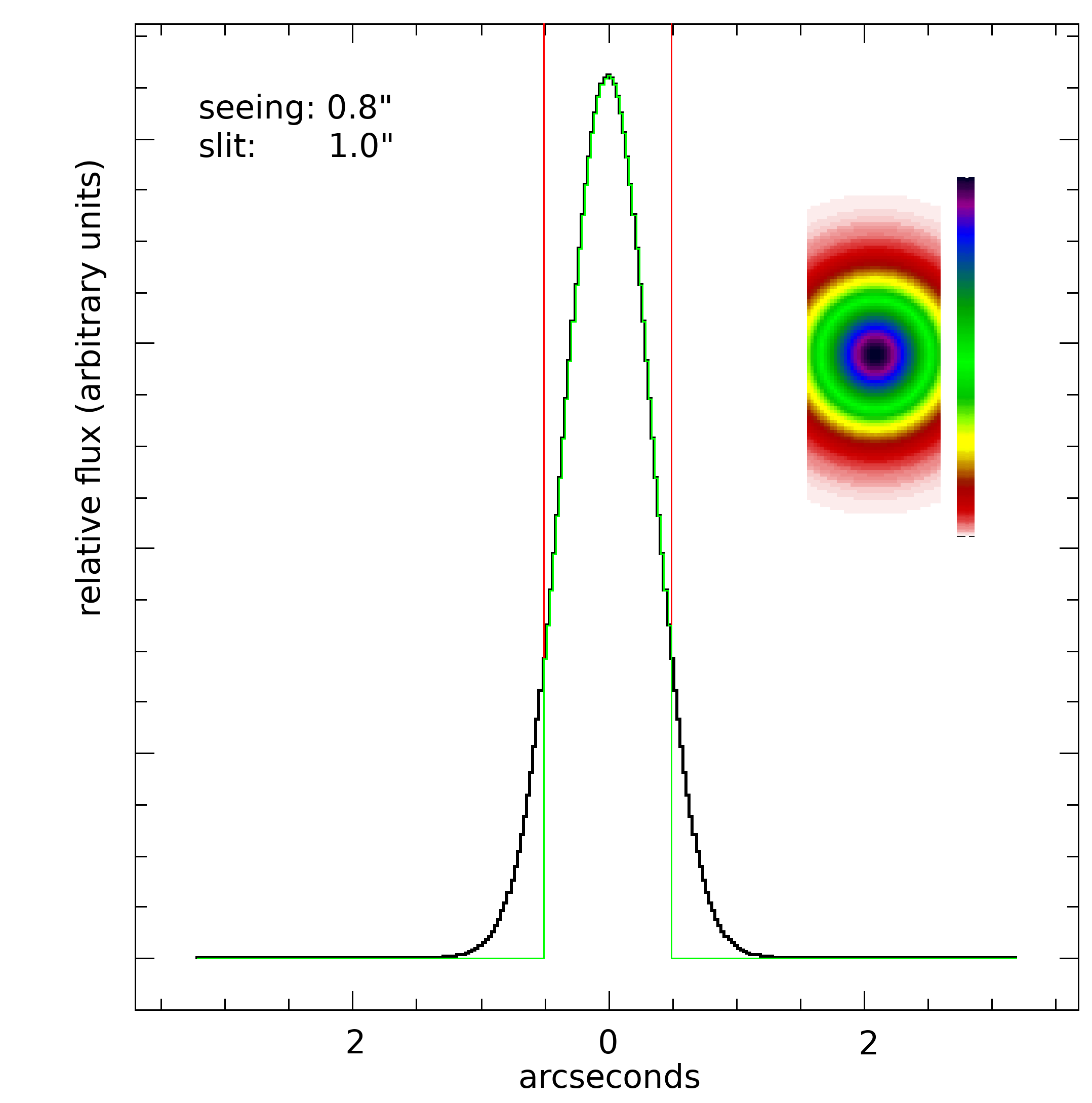}}
\caption{The slit-loss calibration. A cut through the stellar image (black) the slit boundaries (red) and the resulting remaining flux reaching the instrument (green). The insert shows the full 2D representation on a linear color scale.}
 \label{fig:slitloss}
 \end{figure}

 The typical seeing was about 0.7{\arcsec}. However, seeing values up to 1.5\arcsec\ sometimes occurred.  A few spectra had extreme seeing and thus blurred images. The seeing was measured from the FWHM of the stellar images along the slit. The values corresponded well with those given by the ASM. Assuming Gaussian point spread functions, the  slit-losses relative to the median FWHM were derived.  ESO takes the standards for the service mode with very wide slits to be free of slit losses, but the pipeline does not apply a correction for the user's science frames. The flux calibration of a point source with a partially filling PSF in a slit and that of a homogeneously slit-filling extended source differs. For this purpose, a small MIDAS batch was used to derive the factor of the filling by the calibration stars for each frame individually. The fraction of a 2D Gaussian with the derived FWHM was used to obtain the conversion factor (Fig.~\ref{fig:slitloss}). Generally, this was small as well as the slits were normally wider than the seeing.  Figure \ref{fig:histogramms} shows the distribution of the seeing values and the resulting slit efficiencies ($\equiv 1-$slit-loss). A pair of direct consecutive nights with seeing 0\farcs58 and 1\farcs13 were used to estimate the errors of the correction procedure. They are in the order of 0\fm04. The worst images with slit efficiencies below 0.55, which also are those showing blurry spectra, are rejected from the sample. At the PN lines, we assumed, that the illuminations do not vary on the scale of the slit size. As the profiles along the slit also do not show very rapid flux variations  we are confident that this assumption holds. The table containing all correction factors for the whole set of nearly 300 individual spectra can be found in the appendix of \citet{Martin}.
 
\begin{figure}
     \centering
     \includegraphics[width=0.47\textwidth]{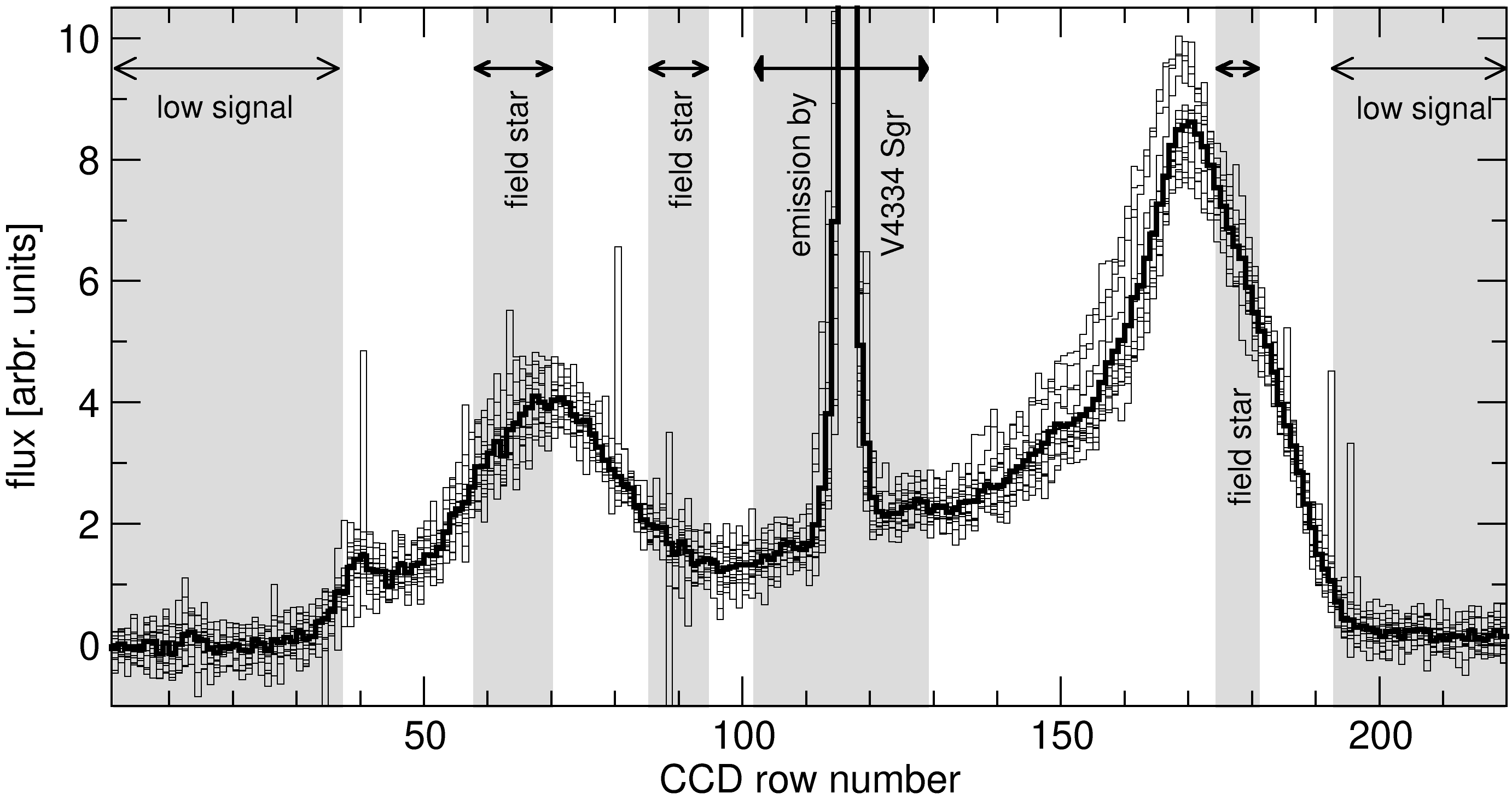}
     \caption{
     The  16 measurements (thin lines) and the derived mean (thick line)  of the [\ion{N}{2}] 6584\,\AA\ line taken in 2010 with the visual band GRISM{s V300 and V600}. The shaded areas are the applied rejection masks 
    %(low signal limit, field star contamination and
    %the region of the emission of the VLTP flash object). 
    The data value $d_i$ is derived as the mean of the integrals from the unmasked regions.}
     \label{fig:mask}
 \end{figure}

\begin{figure}
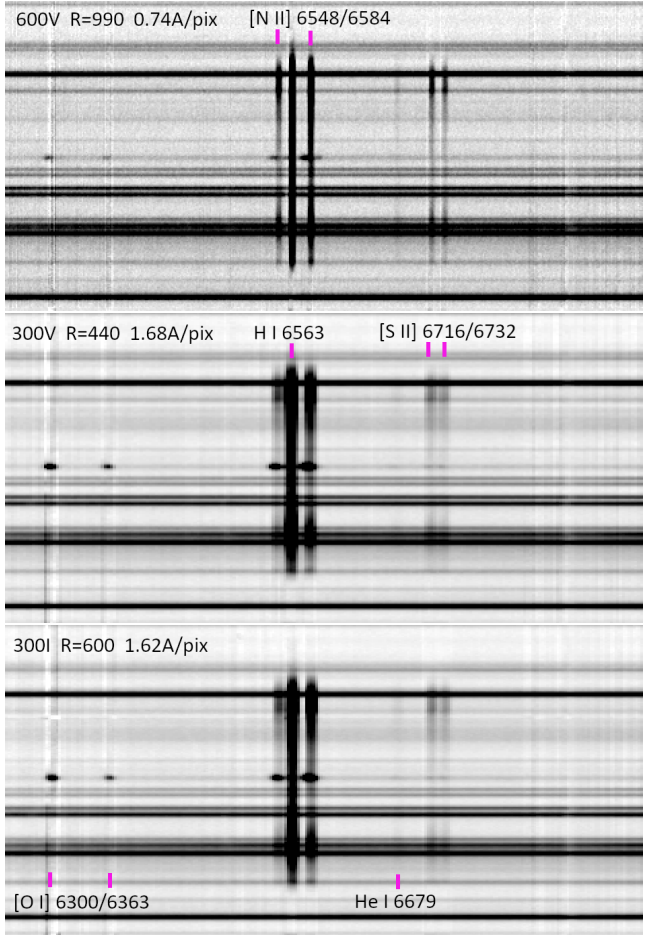

     \centering
     \includegraphics[width=0.47\textwidth]{v600.png}\\
     \includegraphics[width=0.47\textwidth]{v300}\\
     \includegraphics[width=0.47\textwidth]{i300}\\
     \caption{The overlap region of the 600V (upper), the 300V (middle) and 300I (lower) GRISM from 6250\,\AA~ to 6950\,\AA~ after sky emission line removal showing the effects of the different resolutions as well as the contamination by field stars and the clumpy emission of the central object V4334 Sgr itself. }
     \label{fig:spec}
 \end{figure}

 Finally the spectra of each PN line were extracted from the 2D spectra manually by marking interactively the line emission region.
 Several field stars cross the region of the nebula (see Figs. \ref{fig:image} and \ref{fig:spec}) . Normally the stellar flux just beside the emission line is interpolated to correct for that. However, especially around the Balmer lines, the stellar flux under the nebular emission lines may behave differently than predicted by continuum interpolation for the star, due to strong Balmer absorption lines in the stellar spectra. As we are interested in variations only and not in the total PN flux, we masked these regions. Finally, the central region of V4334 Sgr was masked in order to exclude the recent ejecta \citep{Hinkle2014, Hinkle2020}, where the emission lines can be orders of magnitudes brighter than those of the PN \citep{Vanhoof2007, vanhoof2018}. This avoids possible scattered light in the instrument (see Figure~\ref{fig:mask}) arising from those lines of the VLTP core.

Within each set of three spectra of a single OB, the flux of the lines (integrated over the used area along the slit) does not vary beyond the photon noise. The same applies to the rare cases of OBs taken consecutively during the same night. However, between sets taken weeks to months apart, fluxes do vary slightly. As we do not expect physical processes in the PN on those timescales, we assume these variations to originate from the weather and calibration inaccuracies.  These variations give us a final estimate for the errors. For some lines, more than 20 individual measurements are available within some of years. All data points of one line for one GRISM within a year were averaged, giving the data points $d_{i}$ and its  standard deviations $\sigma_i$.
In the case of independent data points, the error of a mean value is smaller than the  standard deviation of the individual measurements. Then, our error would be smaller by a factor of 2 to 4. But we are dominated by systematic and not by statistical errors. Therefore, this approach might be too optimistic.
To use this  standard deviation as error for $d_{i}$ is a very conservative access. The measurements in the overlap region from the visual to the infrared GRISMs are treated as independent data sets. The values used in the statistical analysis are given for each data set and each year in Tab.\ \ref{tab:sigma}.

Sets taken at higher and lower resolution by the 300V/I and 600V/I GRISMs overlap  in the region of the H$\alpha$ line. Figure\ \ref{fig:spec} shows the region of overlap and the various resolution effects. In case of the low resolution GRISMs  the [\ion{N}{2}] 6548\,\AA\ line slightly blends with \ion{H}{1} 6563\,\AA. However, the nitrogen line originates from the same upper excitation level as [\ion{N}{2}] 6584\,\AA. Thus the line ratio is fixed to a value of 2.96 by the ratio of the quantum mechanical transition strengths $A_{ki}$ multiplied by the ratio of the line wavelengths, taken  from the NIST atomic database \citet{NIST}. This  allows us to reconstruct the undisturbed hydrogen flux. The \ion{He}{1} 6679\,\AA\ was too weak to be used for a significant decent fraction of the time series. The [\ion{O}{1}] 6300+6363\,\AA\ pair was significantly affected by the telluric line, and could not be fully recovered. The partly resolved [\ion{S}{2}] 6716+6732\,\AA\ pair was summed to improve signal-to-noise ratio.\\

The data from different instrumental configurations of the same line show small variations between the different sets but do not show systematic trends. Different instrument setups also show no systematic effects. Up to 20 measurements were available per year.  Moreover, the three line pairs \ion{H}{1} 6563+4861\,\AA, [\ion{O}{3}] 4958+5007\,\AA\ and [\ion{S}{3}] 9068+9530\,\AA\ are tightly coupled by quantum mechanics and have well determined line ratios. Their consistency provides an additional quality test for the calibration.

\begin{table*}
    \caption{The measured annual values and the standard deviations normalized to the epoch $t_0 = 2012.45$: $\thickbar{d}_i:=d_{i}/n$ and $\thickbar{\sigma}_i:=\sigma_{i}/n$. The superscript indicates if the values are obtained by the visual (V) band 300V and 600V GRISM  or by an infrared (I) band 300I, 600I and 600z GRISM.
    }
    \label{tab:sigma}
    \centering
    
        \begin{tabular}{l cc cc cc cc cc cccc cccc}
        \hline\hline
year &	\multicolumn{2}{l}{\ion{He}{2} 4686} & \multicolumn{2}{l}{\ion{H}{1} 4861} & \multicolumn{2}{l}{\ion{O}{3} 4958} & \multicolumn{2}{l}{\ion{O}{3} 5007} & \multicolumn{2}{l}{\ion{He}{1} 5875} &
 \multicolumn{4}{l}{\ion{H}{1} 6563} &
 \multicolumn{4}{l}{\ion{N}{2} 6584}\\
     & \!\!$\thickbar{d}_i^{\rm V}$ & \!\!\!\!$\thickbar{\sigma}_i^{\rm V}$	& \!\!$\thickbar{d}_i^{\rm V}$ & \!\!\!\!$\thickbar{\sigma}_i^{\rm V}$	& \!\!$\thickbar{d}_i^{\rm V}$ & \!\!\!\!$\thickbar{\sigma}_i^{\rm V}$	& \!\!$\thickbar{d}_i^{\rm V}$ & \!\!\!\!$\thickbar{\sigma}_i^{\rm V}$
     & \!\!$\thickbar{d}_i^{\rm V}$ & \!\!\!\!$\thickbar{\sigma}_i^{\rm V}$
     & \!\!$\thickbar{d}_i^{\rm V}$ & \!\!\!\!$\thickbar{\sigma}_i^{\rm V}$ & \!\!$\thickbar{d}_i^{\rm I}$ & \!\!\!\!$\thickbar{\sigma}_i^{\rm I}$ 	
     & \!\!$\thickbar{d}_i^{\rm V}$ & \!\!\!\!$\thickbar{\sigma}_i^{\rm V}$ & \!\!$\thickbar{d}_i^{\rm I}$ & \!\!\!\!$\thickbar{\sigma}_i^{\rm I}$ \\	
     \hline
2007&	\!\!0.999&	\!\!\!\!0.436 &  \!\!1.007&	\!\!\!\!0.149&	\!\!1.018&	\!\!\!\!0.078&	\!\!1.016&	\!\!\!\!0.060&\!\!1.128&	\!\!\!\!0.606& 
	\!\!0.979&	\!\!\!\!0.082&	\!\!\!\!&	\!\!\!\!&	\!\!0.831&	\!\!\!\!0.070&	\!\!\!\!&	\!\!\!\!
\\
2008&	\!\!1.007&	\!\!\!\!0.564&\!\!1.073&	\!\!\!\!0.084&	\!\!1.158&	\!\!\!\!0.069&	\!\!1.157&	\!\!\!\!0.074&\!\!1.062&	\!\!\!\!0.380&
	\!\!1.125&	\!\!\!\!0.099&	\!\!\!\!&	\!\!\!\!&	\!\!0.990&	\!\!\!\!0.073&	\!\!\!\!&	\!\!\!\!
\\
2009&	\!\!0.884&	\!\!\!\!0.538&\!\!0.917&	\!\!\!\!0.138&	\!\!0.946&	\!\!\!\!0.067&	\!\!1.006&	\!\!\!\!0.238&\!\!0.770&	\!\!\!\!0.360&
	\!\!0.894&	\!\!\!\!0.056&	\!\!\!\!1.027&	\!\!\!\!0.117&	\!\!0.920&	\!\!\!\!0.088&	\!\!\!\!0.818&	\!\!\!\!0.108
\\
2010&  \!\!1.131&	\!\!\!\!0.879&	\!\!0.964&	\!\!\!\!0.335&	\!\!1.058&	\!\!\!\!0.320&	\!\!1.135&	\!\!\!\!0.381&\!\!0.924&	\!\!\!\!0.696&
	\!\!1.031&	\!\!\!\!0.305&	\!\!\!\!1.027&	\!\!\!\!0.070&	\!\!1.012&	\!\!\!\!0.271&	\!\!\!\!0.924&	\!\!\!\!0.067
\\
2011&&	\!\!\!\!&	    &		&		 &       &       &      \!\!	&\!\!\!\!& 
	\!\!&	\!\!\!\!	0.992&	\!\!\!\!0.070&	\!\!\!\!&	\!\!\!\!&	\!\!0.946&	\!\!\!\!0.098&	\!\!\!\!&	\!\!\!\!
\\
2012&	\!\!1.281&	\!\!\!\!0.807&\!\!1.055&	\!\!\!\!0.110&	\!\!0.974&	\!\!\!\!0.101&	\!\!0.979&	\!\!\!\!0.095&\!\!1.410&	\!\!\!\!1.328&
	\!\!0.994&	\!\!\!\!0.080&	\!\!\!\!1.018&	\!\!\!\!0.074&	\!\!1.017&	\!\!\!\!0.064&	\!\!\!\!1.039&	\!\!\!\!0.123
\\
2013&	\!\!0.937&	\!\!\!\!0.299&\!\!1.102&	\!\!\!\!0.474&	\!\!0.957&	\!\!\!\!0.101&  \!\!1.005&	\!\!\!\!0.108&\!\!1.833&	\!\!\!\!1.663&
	\!\!1.018&	\!\!\!\!0.085&	\!\!\!\!1.146&	\!\!\!\!0.125&	\!\!1.203&	\!\!\!\!0.178&	\!\!\!\!1.035&	\!\!\!\!0.184
\\
2014&	\!\!1.396&	\!\!\!\!1.291&\!\!1.197&	\!\!\!\!0.400&	\!\!1.018&	\!\!\!\!0.104&	\!\!1.027&	\!\!\!\!0.122&\!\!1.391&	\!\!\!\!1.638&
	\!\!1.085&	\!\!\!\!0.098&	\!\!\!\!1.347&	\!\!\!\!0.216&	\!\!1.446&	\!\!\!\!0.196&	\!\!\!\!1.132&	\!\!\!\!0.106
\\
2015&	\!\!1.706&	\!\!\!\!1.474&\!\!1.005&	\!\!\!\!0.065&	\!\!0.906&	\!\!\!\!0.105&	\!\!0.940&	\!\!\!\!0.089&\!\!1.626&	\!\!\!\!1.254&
	\!\!0.983&	\!\!\!\!0.067&	\!\!\!\!1.081&	\!\!\!\!0.150&	\!\!1.163&	\!\!\!\!0.179&	\!\!\!\!1.057&	\!\!\!\!0.063
\\
2016&	& & 	&		&		&        &       &      & & & 
&	&	\!\!\!\!0.972&	\!\!\!\!0.120&	& & \!\!\!\!1.070&	\!\!\!\!0.150
\\
2017&	\!\!0.954&	\!\!\!\!0.815&\!\!0.930&	\!\!\!\!0.073&	\!\!0.862&	\!\!\!\!0.110&	\!\!0.885&	\!\!\!\!0.066&\!\!1.034&	\!\!\!\!0.498&
	\!\!0.904&	\!\!\!\!0.102&	\!\!\!\!0.994&	\!\!\!\!0.158&	\!\!\!\!1.172&	\!\!\!\!0.172&	\!\!\!\!1.092&	\!\!\!\!0.071
\\
2022&	\!\!1.089&	\!\!\!\!0.500&\!\!0.920&	\!\!\!\!0.090&	\!\!0.765&	\!\!\!\!0.130&	\!\!0.774&	\!\!\!\!0.070&\!\!1.002&	\!\!\!\!0.340&
	\!\!0.950&	\!\!\!\!0.070&	\!\!\!\!0.910&	\!\!\!\!0.100&	\!\!\!\!1.325&	\!\!\!\!0.090&	\!\!\!\!1.250&	\!\!\!\!0.130
\\
\hline
year &	\multicolumn{4}{l}{\ion{S}{2} 6716+6732} &
\multicolumn{4}{l}{\ion{Ar}{3} 7136} & 
\multicolumn{2}{l}{\ion{S}{3} 9068} &
\multicolumn{2}{l}{\ion{S}{3} 9560} 
\\
& \!\!$\thickbar{d}_i^{\rm V}$ & \!\!\!\!$\thickbar{\sigma}_i^{\rm V}$ & \!\!$\thickbar{d}_i^{\rm I}$ & \!\!\!\!$\thickbar{\sigma}_i^{\rm I}$ &
\!\!$\thickbar{d}_i^{\rm V}$ & \!\!\!\!$\thickbar{\sigma}_i^{\rm V}$ &\!\!$\thickbar{d}_i^{\rm I}$ & \!\!\!\!$\thickbar{\sigma}_i^{\rm I}$ &
\!\!$\thickbar{d}_i^{\rm I}$ & \!\!\!\!$\thickbar{\sigma}_i^{\rm I}$ & \!\!$\thickbar{d}_i^{\rm I}$ & \!\!\!\!$\thickbar{\sigma}_i^{\rm I}$ \\
\hline
2007 & 0.994& \!\!\!\!0.161& \!\!\!\!& \!\!\!\!	&	\!\!0.681& \!\!\!\!0.429		\\				
2008 & 1.160& \!\!\!\!0.229& \!\!\!\!& \!\!\!\!	&	\!\!1.015& \!\!\!\!0.097& \!\!\!\!0.941& \!\!\!\!0.157\\
2009 & 0.959& \!\!\!\!0.339& \!\!\!\!0.771& \!\!\!\!0.224&	\!\!0.866& \!\!\!\!0.371& \!\!\!\!0.940& \!\!\!\!0.770	&	\!\!0.959& \!\!\!\!0.107	&	\!\!0.954& \!\!\!\!0.096\\
2010 & 0.865& \!\!\!\!0.179& \!\!\!\!0.873& \!\!\!\!0.098&	\!\!1.133& \!\!\!\!0.604& \!\!\!\!0.950& \!\!\!\!0.798	&	\!\!0.921& \!\!\!\!0.087	&	\!\!0.942& \!\!\!\!0.100\\
2011 &	& \!\!\!\!& \!\!\!\!0.932& \!\!\!\!0.115		&	\!\!0.887& \!\!\!\!0.272	&	\!\!1.072& \!\!\!\!0.192	&	\!\!1.044& \!\!\!\!0.107\\
1012 & 1.015& \!\!\!\!0.164& \!\!\!\!1.139& \!\!\!\!0.276&	\!\!1.086& \!\!\!\!0.377& \!\!\!\!1.381& \!\!\!\!0.728	&	\!\!1.048& \!\!\!\!0.081	&	\!\!1.043& \!\!\!\!0.135\\
2013 & 0.994& \!\!\!\!0.227& \!\!\!\!1.193& \!\!\!\!0.276&	\!\!1.057& \!\!\!\!0.706& \!\!\!\!0.825& \!\!\!\!0.558	&	\!\!1.100& \!\!\!\!0.165	&	\!\!1.093& \!\!\!\!0.176\\
2014 & 1.526& \!\!\!\!0.264& \!\!\!\!1.099& \!\!\!\!0.199&	\!\!1.352& \!\!\!\!0.772& \!\!\!\!1.486& \!\!\!\!1.518	&	\!\!1.292& \!\!\!\!0.261	&	\!\!1.234& \!\!\!\!0.210\\
2015 & 1.249& \!\!\!\!0.405& \!\!\!\!1.093& \!\!\!\!0.230&	\!\!1.162& \!\!\!\!0.635& \!\!\!\!1.219& \!\!\!\!0.353	&	\!\!1.128& \!\!\!\!0.205	&	\!\!1.015& \!\!\!\!0.124\\
2016 & 1.193& \!\!\!\!0.316& \!\!\!\!& \!\!\!\!1.305	&	\!\!0.902& \!\!\!\!& \!\!\!\!		&	\!\!0.884& \!\!\!\!0.116	&	\!\!0.955& \!\!\!\!0.161\\
2017 & 1.128& \!\!\!\!0.175& \!\!\!\!1.174& \!\!\!\!0.555&	\!\!0.918& \!\!\!\!0.476& \!\!\!\!1.162& \!\!\!\!0.878	&	\!\!1.081& \!\!\!\!0.309	&	\!\!0.947& \!\!\!\!0.306\\
2022 & 1.273& \!\!\!\!0.231& \!\!\!\!1.340& \!\!\!\!0.250&	\!\!1.100& \!\!\!\!0.380			&	\!\!1.196& \!\!\!\!0.104	&	\!\!1.134& \!\!\!\!0.114\\
\hline
    \end{tabular}
\end{table*}

\newpage
\section{Results}% and Conclusions}

\begin{figure}[t]
     \centering
     \includegraphics[width=0.47\textwidth]{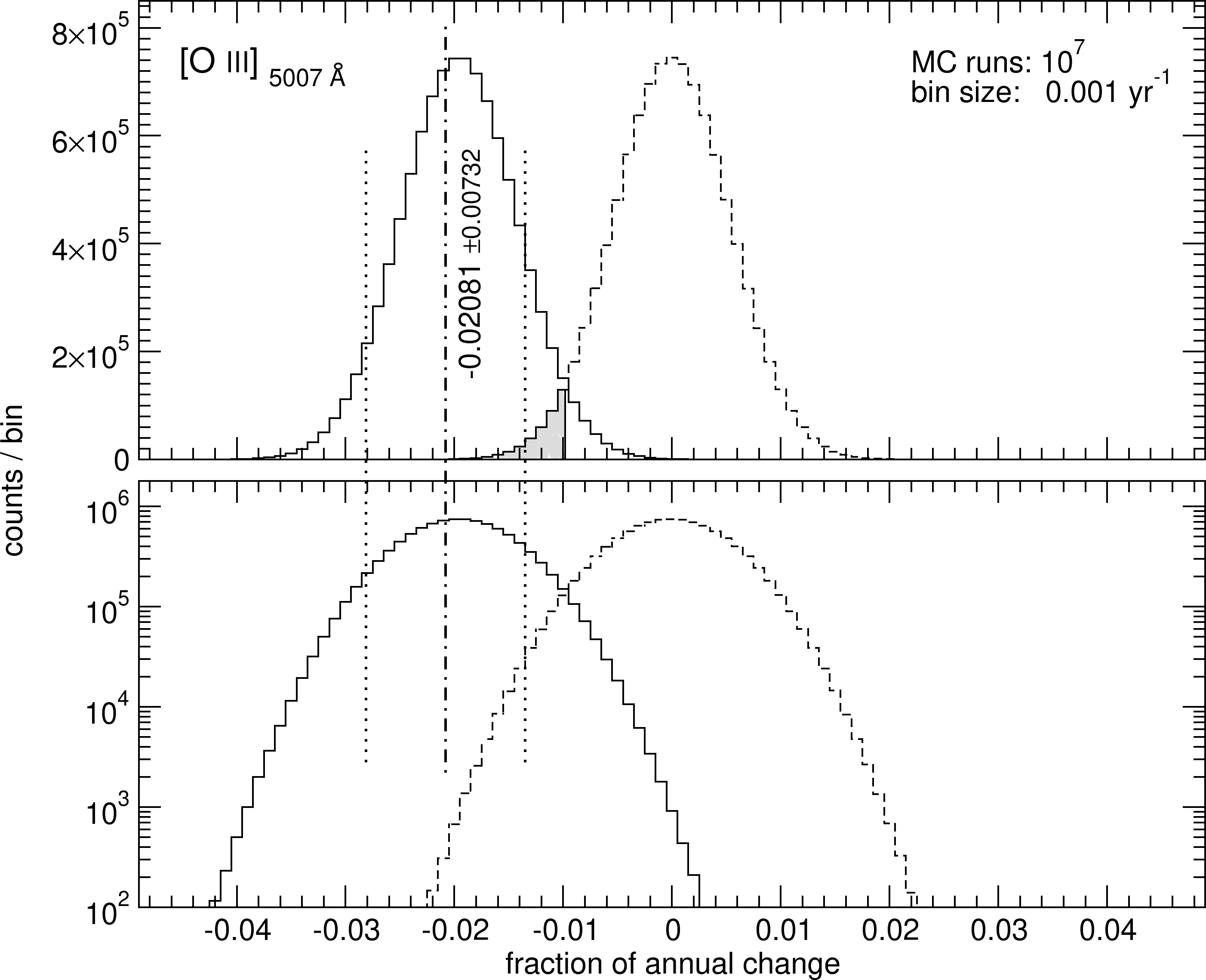}
     \caption{The distributions of $C_{k}$ parameters (solid line) and of the constant solution (dashed line) were derived using the MC runs for the [\ion{O}{3}] line at 5007\,\AA. The shaded area gives the parameter $A$ for the determination of $x$ in equation~\ref{equ:05}. The fit and the region of the standard error derived from a normal regression program are shown for reference with the dash-dotted and dotted lines. }
     \label{fig:hist}
 \end{figure}

We analyzed the change as a function of time, by using  annual groups of line intensities for each instrumental setup as independent data points $d_{i}$. The baseline is taken as time $t_{0} = 2012.45$, which is the mean time point of our original data,  16.31 years after the discovery of the event \citep{Nakano1996}. \ The total process follows an exponential decline (see Equ.\ \ref{equ:07}). As the total timescale $\tau_{rec}$ is so much longer than the epoch of our investigation, a free parameter, giving the curvature of the exponential, cannot be derived unambiguously and numerically stable. Thus we use the Taylor series linear approximation here. We derive  independent model regression points $m_{i}$ of the type
\begin{equation}
\begin{split}
m_{i} = & \,c\,\,(t_{i} - t_0) + n \\
 \hbox{where}\quad & \sum_i {\frac{(d_{i}- m_{i})^2}{\sigma_{i}^2}} \rightarrow {\rm Min}\\
 \end{split}
    \label{equ:02}
\end{equation}
{ with $c$ being the average annual change.} As mentioned above the errors $\sigma_{i}$ vary strongly for some lines (see also Tab.\,\ref{tab:sigma}). Thus the Equ.\,\ref{equ:02} does not resemble the $\chi^2$ definition where $\sigma_i \propto \sqrt{m_i}$ anymore.
Thus standard regression \ algorithms used widely\ do not apply \citep{York66,LastSquares2,LastSquares3}. The regression analysis follows  \citet{Tell2020}.
The derived value of $n$  normalizes each of the data sets with respect to the line strengths at $t(0)$. This is slightly different from using just a weighted mean for the time $t_0$, but is numerically more stable against the strong year-on-year variations of the errors  which are seen especially for the helium lines.
For the regression the {\tt mpfit} library \citep{MPFIT} was used, with the variable errors handled according to \citet{York66} in the implementation by \citet{Tell2020}.
 The errors given by the {\tt mpfit} library for the $c$ parameter represent the statistical error with two parameters for the calculation of the degree of freedom. However, we are primarily interested in the significance of any slope, rather than the parameter $c$ (respectively its normalized counterparts $C_{k}$ from equation~\ref{equ:03}) and its potential contribution to the statistical error budget. To derive the significance of the slope, a {\tt C} program was written to perform a Monte Carlo (MC) simulation. { The value $n$ from Equ. \ref{equ:02} defines the normalized values $\thickbar{d}_i:=d_{i}/n$ and $\thickbar{\sigma}_i:=\sigma_{i}/n$ given in Tab.\ \ref{tab:sigma}.} Each data point $\thickbar{d}_{i}$ was varied independently 10 million times in agreement with its individual Gaussian error distribution and a new regression for the normalized change $C_{k} \,\,\forall \,k \in [1,10^{7}]$ was calculated with  model points $M_{i}$ types
\begin{equation}\label{equ:03}
\begin{split}
M_{i} = & \, C_{k}\,\,(t_{i} - t_0) + N_{i} \\
 \hbox{where}\quad & \sum_i {\frac{(\thickbar{d}_i - M_{i})^2}{\thickbar{\sigma}_i^2}} \rightarrow {\rm Min.} \\
 \end{split}
\end{equation}
 Moreover, a similar set of parameters from the same MC simulated data points, assuming that there was no change in time-defining model points $M$, was derived.
\begin{equation}
\begin{split}
M_{i} = & \, N_{i} \qquad\\
 \hbox{where}\quad&\sum_i {\frac{(\thickbar{d}_i - M_{i})^2}{\thickbar{\sigma}_i^2}} \rightarrow {\rm Min.}\\
 \end{split}
    \label{equ:04}
\end{equation}
The fractional area of overlap $A$ between the 'sloping' and 'non-sloping' distribution functions yields the statistical significance ($1-A$) of the slope as a single parameter. This significance is lower than what would be derived from 
%, due to the different degree of freedom, thus is also lower than 
the standard deviation of the inclination (relative change per year) given by the fit with two free parameters. 
Figure \ref{fig:hist} shows an example of such a pair of histograms. A-priory this solution of the MC simulation, caused by the wide spread of errors between the individual data points, does not have to be distributed Gaussian. Tests indeed showed that it deviates from solutions with large slopes. As the result for our cases only shows very small slopes, there is only a marginal deviation from Gaussian. As the assumption of a regular standard deviation holds we are able to use the error function {\rm erf} from the integral of the Gaussian according to equation~\ref{equ:05} to derive the size of the intersect, 

\begin{equation}
A = 1 - {\rm erf}{\left(- {\frac{x}{\sqrt{2}}}\right)}
    \label{equ:05}
\end{equation}

\noindent and from this the level $x \times \sigma$ of significance.

\begin{figure}[t]
\includegraphics[width=0.47\textwidth]{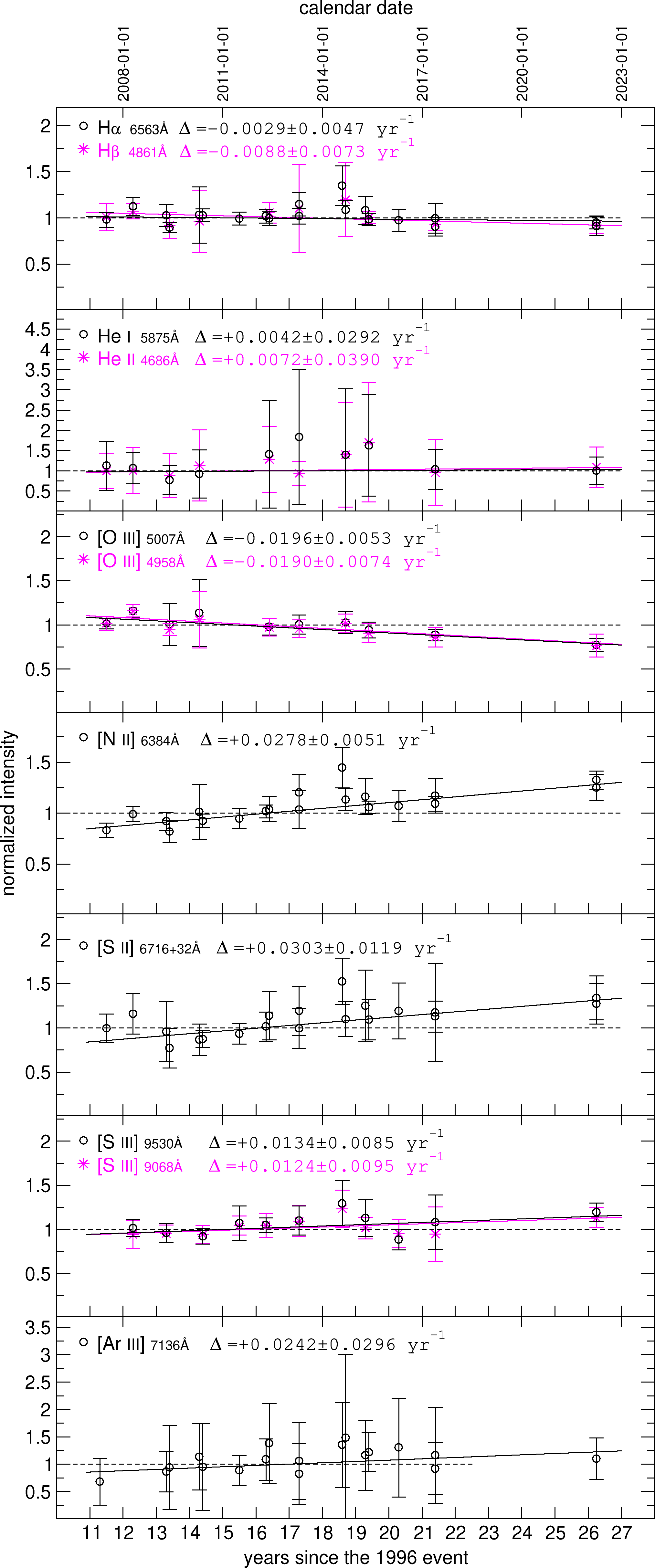}
\caption{The time evolution of various permitted lines of \ion{H}{1}, \ion{He}{1} \& \ion{He}{2} and the forbidden line transitions of [\ion{O}{3}], [\ion{N}{2}], [\ion{S}{2}] \& [\ion{S}{3}] and [\ion{Ar}{3}]. The best linear fit is indicated. The dashed lines indicate the constant solution.}\label{fig:result}
\end{figure}

\begin{table}[t]
    \caption{The annual changes of the various lines together with the significance $x$ in terms of $\sigma$ relative to a non-changing calculation in the MC simulations. The only solutions with  $x> 1.5\,\sigma$ are underlined.
    }
    \label{tab:result}
    \centering
    
        \begin{tabular}{l c c c l}
        \hline\hline
      & Wavelength & $A$     & $x$     & Relative change   \\
    Ion & [\AA] & & $\times \sigma$ & \phantom{rel}per year $\Delta$ \\
    \hline
    \ion{H}{1} & 6563 & 0.465 & 0.73 & $-0.0029 \pm 0.0047$ \\
    \ion{H}{1} & 4861 & 0.325 & 0.98 &$-0.0088\pm0.0072$  \\
    \ion{He}{1} & 5875 & 0.493 & 0.69 & \ \ \,$0.0042\pm0.0292$ \\
    \ion{He}{2} & 4686 & 0.558 & 0.59 &\ \ \,$0.0072\pm0.0390$ \\
    \ion{N}{2}  & 6584 & 0.0076 & \underline{2.66}&\ \ \,$0.0278\pm0.0051$\\
    \ion{O}{3} & 4958 & 0.1255 & \underline{1.53} & $-0.0190\pm0.0074$ \\
    \ion{O}{3} & 5007 & 0.0447 & \underline{2.01} & $-0.0196\pm0.0053$ \\
    \ion{Ar}{3} & 7136 & 0.349 & 0.94 & \ \ \,$0.0242\pm0.0296$ \\
    \ion{S}{2} & 6716+32& 0.112 & \underline{1.59} & \ \ \,$0.0303\pm0.0119$ \\
    \ion{S}{3} & 9068 & 0.283 & 1.07 &\ \ \,$0.0124\pm0.0095$ \\
    \ion{S}{3} & 9530 & 0.260 & 1.13 & \ \ \,$0.0134\pm0.0085$ \\
    \hline
    \end{tabular}
\end{table}

\ Despite the fact, that some lines are coupled by physical and quantum mechanical processes, each line was treated 
individually in this statistical process. This gives additional confidence about the stability of the analysis. The results are shown in Figure\ \ref{fig:result} and Table\ \ref{tab:result} lists the annual changes and the statistical uncertainties compared to a constant solution derived from the MC simulations. 
The hydrogen and the helium lines show, as expected by most models, no variations with time during our survey. Both [\ion{O}{3}] lines show independently the same weak decline in intensity by about 2 percent per year, while the low ionized species [\ion{N}{2}] and [\ion{S}{2}] show a weak strengthening of about 3 percent per year. The effects are small compared to the errors just at the boundary to be detected and thus support the models predicting a slow evolution  (see discussion below). Moreover, the identical behavior of the independently treated pairs of [\ion{O}{3}] and of [\ion{S}{3}] strengthens the statement of the evolution found.

\section{Discussion} 
The $Z^2$ dependence of the recombination rate on the charge of an ion (see \mbox{Eqn. \ref{equ:01})} causes  the observed states of O$^{++}$, S$^{++}$ and Ar$^{++}$ to react fastest to the sudden loss of the UV radiation field. As the observed line levels are collisionally excited, the line strength is related to the total population of the species for constant electron temperature. E.g., the [\ion{O}{3}] lines in our spectra originate from the first exited term above the ground state and are mostly collisionally excited. The line intensity of the 5007/4958\,\,\AA\ transitions thus is directly proportional to the amount of O$^{++}$.

The population of the particular ionization state decreases because of direct recombination losses and increases because of recombination of the higher ionization states (O$^{\rm 3+}$, S$^{\rm 3+}$ and Ar$^{\rm 3+}$).  CLOUDY photoionization models of the pre-outburst PN  by \citet{Pollacco99, Pollacco02, Juffinger2021} show that O$^{\rm 3+}$ was not populated significantly ($\leq 3$\%), while S$^{\rm 3+}$ and Ar$^{\rm 3+}$ had a population of about 8\% each. Hence, recombination for the higher levels can be ignored for these three species. However, more detailed estimates follow below. %only O$^{++}$ can be discussed looking isolated to the timescale of the recombination. 

The fractional population of various ionization stages are listed in Table \ref{tab:ionization}. These are calculated from the CLOUDY model \citet{Juffinger2021}, and averaged over an area to account for the narrow slit used in the observations.

\begin{table}[h]
    \caption{The fractional population of selected ionization stages from the CLOUDY model of \citet{Juffinger2021}, in percent
    }
    \label{tab:ionization}
    \centering
    
        \begin{tabular}{l c c c l}
        \hline\hline
 Ionization stage &           [II] & [III] & [IV] \\

 Helium      &  81 &  18 \\
 Carbon      &  15 &  80 &  \phantom{0}4 \\
 Nitrogen   &  27 & 67 &  \phantom{0}6 \\ %
 Oxygen     &  26 &  70 &  \phantom{0}3 \\
 Sulphur    &  18 &  73 &  \phantom{0}8 \\
 Chlorine   &  14 &  76 &  10 \\
 Argon      &  \phantom{0}3 &  90 &  \phantom{0}7 \\  
     \hline
    \end{tabular}
\end{table} 

For the radiative recombination factor $\alpha_{\rm RR}(T,Z)$ we used the data from  \citep{alpha_rr_2006}, while for the dielectric component $\alpha_{DR}(T,Z)$ the calculation is based on the C-like sequence of \citet{alpha_dr_2004} in the updated web-based version of the group \citep{BadnellWeb}. The same data is used for the current version of CLOUDY c17 \citep{Cloudy2017} as well. Following \citet{OF06} these two components can be simply added together to obtain $\alpha(T,Z)$. The modern values differ slightly from the often used values in the tables of \citet{OF06}, which are based on somewhat older calculations \citep[][and references therein]{Verner1996,Mazzotta1998}. The older values were used to estimate timescales by \citet{Pollacco99, Lechner2004}, and lead to a somewhat faster than expected evolution.

The recombination timescale $\tau_{\rm rec}$ is then given by
\begin{equation} \label{equ:06}
\tau_{\rm rec} = \frac{1}{\alpha(Z,T_{\rm e}) \,\,n_{\rm e}}
\end{equation}
with $Z$ the ionic charge, $T_{\rm e}$ the electron temperature and $n_{\rm e}$ the electron density. Diagnostic diagrams of observed spectral line ratios yield  $T_{\rm e} \gtrapprox 11\,000\,{\rm K}$ and  $140 \le n_{\rm e} \le 180\,{\rm cm}^{-3}$  \citep{Pollacco99,Kerber1999,Juffinger2021}. These values are typical for old PNe generally \citep[see e.g. ][]{oettl2014, Barria2018}. This density is much lower than the one used in the test for dynamical evolution with CLOUDY by \citet{Cloudy2020}. We use the upper limit of $n_{\rm e}\,\approx\,180\,{\rm cm}^{-3}$ for our estimates. Lower values will result in an even slower evolution.

The timescale for the recombination of the small (3\%) O$^{\rm 3+}$ population becomes only 13 years. Accounting for this and assuming all  O$^{\rm 3+}$ was added quickly to  O$^{++}$, this ionization state covers more than 73\% of the total amount of oxygen in the CLOUDY models.  The predicted timescale of the O$^{++}$ recombination $t \approx\,\,61$~years leading to about 25\% of decline from the 1996 outburst until the epoch at the end of our observations. 

This prediction is within the bounds of the observational result in Tab.~\ref{tab:result}. Using an undisturbed and thus exponential decline,
\begin{equation} \label{equ:07}
\begin{split}
n_{\rm ion}(t) = n_{\rm ion}(0) \exp\left({-\,\frac{t}{\tau_{\rm rec}}}\right)\phantom{.} \\ 
\frac{\rm d}{{\rm d}t} n_{\rm ion}(t) = -\, n_{\rm ion}(0) \frac{1}{\tau_{\rm rec}} \exp\left({-\,\frac{t}{\tau_{\rm rec}}}\right), \\ 
 \end{split}
\end{equation}
leads in our observations to a timescale of about 50 years. Within the uncertainties of the density estimates and the statistical uncertainties this matches well. The rate of decline is about an order of magnitude slower than the values given in \citet{Pollacco99}.

The population of S$^{\rm 3+}$ and Ar$^{\rm 3+}$ with timescales of 11 and 16 years respectively, would depopulate those states  by about 75\% each.
This adds about 7\% to  S$^{++}$ and Ar$^{++}$. The latter states, originally populated by about 73 and 90\% of the total sulfur and argon, have timescales of 40 and 50 years to recombine to S$^{+}$ and Ar$^{+}$. Thus even with the compensation by the faster recombining of higher levels, we predict a decline of about 50\%. The observations do not confirm a significant decline. Both states show an increase at the 1$\sigma$ level. Due to the higher errors, the Ar$^{++}$ observations are consistent with the predicted time scale only at the $2\sigma$ level (probability $\lessapprox 2$\%). For S$^{++}$, the discrepancy is in excess of 4.5\,$\sigma$ and thus highly significant.

The observed increase of the lines of the low ionization stages N$^{+}$ and S$^{+}$ reflect their populations of respectively 27 and 18\% pre-outburst. Their lifetimes are 250 years or more. Thus even a limited fraction of recombination from the upper ionization states will yield a significant strengthening of the  [\ion{N}{2}] and [\ion{S}{2}] lines.

The recombination timescales calculated for hydrogen H$^{+}$ and He$^{+}$ are 450 years or more. This is supported well by the observations, which do not indicate any changes in these emission lines. Moreover, as those elements outnumber the other ions by two orders of magnitude, the fraction of the total energy density bound in ionization does not change at all.

The nebular expansion could also affect flux evolution due to decreasing electron density. However, the measured expansion of 30$\pm$3 km s$^{-1}$ \citep{Juffinger2021} leads to an estimated decrease by a factor of $7 \times 10^{-4}$ per year and unit volume. This is much slower than the expected effect from recombination and below the accuracy of the observations themselves. 

\section{Sulfur}

A notable discrepancy with the predicted evolution is seen for sulfur, where the [\ion{S}{3}] lines are constant in strength whereas the models predict a decrease. This can  come from an additional source of S$^{2+}$ which balances the recombination. 

Measured sulfur abundances in planetary nebulae are anomalously low \citep{Henry2012} compared to other elements. The sulfur deficit is up to a factor of 2. Possible causes are discussed in \citet{Henry2012}. They state that in some cases, telluric absorption near the [\ion{S}{3}] 9068+9530\,\AA~may be present. In our case, the resolution is very good and the very long FORS slits allow excellent sky reduction. However, the sulfur problem is most pronounced for high ionization, matter-bounded nebulae. This suggests that the models may be missing higher ionization  stages, leading to underestimated ionization correction factors. Direct measurements of infrared [\ion{S}{4}] lines have provided support for this conclusion \citep{Henry2012}, as these find a higher than predicted abundance of this ionization stage for these nebulae.

The old PN of Sakurai's star falls in the category of high excitation, matter-bounded nebulae. If the S$^{3+}$ abundance is higher than predicted in the model, the fast recombination from this level would increase the source function of S$^{2+}$. Higher ionization stages, S$^{4+}$ and higher, could contribute since they would recombine very fast to S$^{3+}$.

We recalculated the expected abundance change of S$^{2+}$ for variable abundances of S$^{3+}$ and higher ionization stages. The measured change with $1-\sigma$ errors can be reproduced if the higher ionization stages account for $65\pm16$\%\ of sulfur, as opposed to 10\%\ in the CLOUDY model. This gives a total sulfur abundance that is around 55\%\ higher than in the model, which is within the typical range of the sulfur anomaly.

A sulfuric dust component, which is another explanation for the anomaly, would not affect the recombination rates. More accurate data will be needed for this analysis, but the apparent recombination discrepancy suggests, that higher ionization stages may be involved in part or all of the observed sulfur anomaly. 

\section{Conclusions}

We were able for the first time to follow the recombination evolution of a very low-density galactic gaseous nebula. careful analysis determines the order of magnitude of the effects of recombination. This result excludes by orders of magnitudes some faster evolutionary scenarios postulated before. 
While oxygen seems to follow the predicted evolution for updated atomic data, other metals do not agree as well, and especially sulfur deviates significantly from the predictions.  

For sulfur, the lack of fading of the [\ion{S}{4}] lines points to possible recombination from unobserved higher ionization levels. There is some evidence that the 'sulfur' anomaly (a sulfur abundance deficit) may be caused by unexpectedly high levels of S$^{4+}$. The current investigation has led to some support for this.  Other explanations have also been proposed and it is important to continue these studies.

The high efficiency of energy transfer by fluorescence calculated by \citet{Deguchi} suggests that this increases the ionization levels of the metals.   Electron transfer reactions from the long-lived  ionized hydrogen and helium may also add to this. As those long-living species outnumber the metals by orders of magnitudes, their excitation energy and electron losses could affect the metals without significantly changing their own abundances. %Moreover, the excitation by various processes, as shown by \citet{Deguchi}, as recombination from exited states is more difficult, may further change these rates. %by the already mentioned Bowen florescence with the He$^+$ will cause highly excited states of the O$^{++}$. 
%As recombination from an excited state is more difficult than that from the ground state, 
The theoretical recombination rates may therefore be subject to change. 

In conclusion, these observational results can be useful for validating more complex time-evolution models \citep[see e.g.][and references therein]{Cloudy2020} which are in development. The simplified recombination timescale calculated previously \citep{Pollacco99, Lechner2004, Balick21} may have to be revisited, while   the estimates of an even much faster evolution by \citep{Schoenberner08}, assuming timescales well below a year, are not confirmed. Further monitoring of V4334 Sgr will be needed to improve the significance  of the result but the results are encouraging and unique.

%Only once we are able to model this unique target, the extrapolation to the older twin V605 Aql as provided by \citet{Lechner2004} has to be revisited. Thereafter, we can go for the recombination in mixed systems without complete loss of the UV radiation like the LTP stars \citep{SwSt1, Balick21}, the very outskirts of old evolved PNe showing recombination halos \citep{vanHoof2000,Corradi2000,Corradi2003} or the old post common envelopes of BE~UMa stars \citep{YY_Hya2020}, suffering partly from re-ionization. However, in none of those cases we have such a good coverage in time for the observational data.

%\begin{figure*}
%\includegraphics[width=60mm]{plot_H_N.png}
%\includegraphics[width=60mm]{plot_He_Ar_N.png}
%\includegraphics[width=60mm]{plot_S_O_S.png}
%\caption{The time evolution}\label{evolutions}
%\end{figure*}

%\hfill

%\phantom{X}
%\newpage

%\begin{acknowledgments}
\bigskip

\noindent
{\scriptsize This investigation makes use of ESO data from program IDs \mbox{077.D-0394}, \mbox{079.D-0256}, \mbox{381.D-0117}, \mbox{383.D-0427}, \mbox{385.D-0292}, \mbox{087.D-0223}, \mbox{089.D-0080}, \mbox{091.D-0209}, \mbox{093.D-0195}, \mbox{095.D-0113}, \mbox{097.D-0146} \mbox{099.D-0045}, and \mbox{0109.D-0060}. This research has made use of the SIMBAD database, operated at CDS, Strasbourg, France, and of the NASA's Astrophysics Data System. MH acknowledges financial support from National Science Centre, Poland, grant No. 2016/23/B/ST9/01653. AAZ acknowledges support from STFC under grant ST/T000414/1.
%\end{acknowledgments}


%%%%\vspace{5mm}
\vspace{3mm}
%\facilities{
\noindent{\it Facilities:} VLT:Antu, Kueyen (FORS1/2)%}

%\software{
\noindent{\it Software:} ESO MIDAS \citep{MIDAS_1,MIDAS_2}%}
}
          
\bibliography{biblio}{}
\bibliographystyle{aasjournal}
          
%\begin{appendix}
%\section{Tables (electronic version only)}

%\input{tab_obslog.tex}
%\input{tab_linefluxes.tex}
%\end{appendix}
\end{document}